# Alternative formulas for synthetic dual system estimation in the 2000 census

## Lawrence Brown[*1] and Zhanyun Zhao[†2]

*University of Pennsylvania and Mathematica Policy Research*

**Abstract:** The U.S. Census Bureau provides an estimate of the true population as a supplement to the basic census numbers. This estimate is constructed from data in a post-censal survey. The overall procedure is referred to as dual system estimation. Dual system estimation is designed to produce revised estimates at all levels of geography, via a synthetic estimation procedure.

We design three alternative formulas for dual system estimation and investigate the differences in area estimates produced as a result of using those formulas. The primary target of this exercise is to better understand the nature of the homogeneity assumptions involved in dual system estimation and their consequences when used for the enumeration data that occurs in an actual large scale application like the Census. (Assumptions of this nature are sometimes collectively referred to as the "synthetic assumption" for dual system estimation.)

The specific focus of our study is the treatment of the category of census counts referred to as imputations in dual system estimation. Our results show the degree to which varying treatment of these imputation counts can result in differences in population estimates for local areas such as states or counties.

## 1. Introduction

The U.S. census is required by the Constitution to be conducted every ten years. In an attempt to provide better estimates of the true population than contained in the basic census counts, the Census Bureau [13] uses both statistical and demographic methods. In 2000 the statistical process was called Accuracy and Coverage Evaluation (A.C.E.).

The 2000 A.C.E. data consists of two parts: the Population sample (P-sample) and the Enumeration sample (E-sample). The P-sample includes persons who are validly included in the A.C.E. survey, and the E-sample includes census enumerations from households in the A.C.E. block clusters. For a detailed overview of the 2000 A.C.E., please see Hogan [9] and Norwood and Citro [11].

The 2000 A.C.E. was designed to get an estimate of the population at every geographic level, based on the census count and the information from the E-sample and the P-sample. To be more precise, the procedure adopted by the Census Bureau is termed a synthetic dual system estimate. Its validity rests on several assumptions, including a major synthetic (homogeneity) assumption.

---

[*]Supported in part by NSF Grant DMS-99-71751.

[†]Supported in part by National Research Council of the National Academy of Sciences.

[1]University of Pennsylvania, 400 Jon M. Huntsman Hall, 3730 Walnut Street, Philadelphia, PA 19104, USA, e-mail: lbrown@wharton.upenn.edu

[2]Mathematica Policy Research, Inc., P.O. Box 2393, Princeton, NJ 08543, USA, e-mail: zzhao@mathematica-mpr.com

Various technical assumptions can be made for synthetic assumption. These affect the details of the formulas needed to produce the final population estimates. For ideal and homogeneous populations any of the resulting formulas will produce unbiased estimates. However, the U.S. population does not appear to have this type of ideal structure. Hence different synthetic assumptions yield different estimates, and it does not appear that all of these estimates are actually unbiased.

This paper investigates the nature of these assumptions and the extent of the differences produced when using three alternative dual system formulas within the 2000 U.S. Census. It should be emphasized that the data available to us do not allow us to make any confident claim as to which of the estimates is more accurate; indeed such a claim is not our objective. Instead, we present our analyses as a means of providing better understanding of the dual system estimation process in the presence of actual populations, such as that encountered in the 2000 Census, and of judging the extent of differences that may be expected to result from differing assumptions about the census enumeration process.

Our analysis revolves around the extent and homogeneity of imputations of household and whole person records into the census enumeration. The available data allows us to produce alternative estimates based on different treatment of these imputations. As we later remark, there are other aspects of the dual system process that might involve analogous biases in the presence of inhomogeneity, however the data available to us do not allow for as complete an analysis relative to those factors.

In Section 2 we briefly discuss the nature and extent of imputation in the 2000 census. It is clear that the desired stochastic homogeneity does not hold there. Section 3 introduces background for dual system estimation and the synthetic assumption. The alternative formulas are presented in Section 4. Section 5 displays the results of using these formulas to estimate the true population shares of the states in 2000. Section 6 presents similar results for estimation of population shares of groups of counties. Mathematical comparison of different formulas is made in Section 7. Section 8 contains a summary conclusion and remarks.

The data for A.C.E. was collected during the 2000 census and first prepared and analyzed before April 2001. The Census Bureau decided not to issue the results then produced as official census estimates. Following this, the data was re-analyzed several times, leading up to revised A.C.E. estimates, referred to as A.C.E. Revision II. These were released on March 2003. The revised data identified, and deleted from the estimation process, a significant number of records that were judged to be duplicates. There were also a number of other more technical, but not insignificant, innovations in A.C.E. Revision II. See Kostanich [10] for a more complete description of A.C.E. Revision II.

The analyses of our paper are based on the original April 2001 A.C.E. data. There are several reasons for our using this original data, rather than the revised A.C.E. II data. The primary reason is that this is the data that was supplied to us by the Bureau, beginning in 2001. (We gratefully acknowledge the Bureau's assistance in supplying us with suitable versions of this data.) Furthermore, our purpose has been to understand the nature of traditional dual system estimation, and the consequences of alternate synthetic assumptions. For the most part the nature of the April 2001 A.C.E. data in relation to the census is analogous to that between earlier censuses and their dual system surveys. (In particular, both the 2000 census counts and the 2001 A.C.E. data contain correspondingly significant numbers of duplicates, such as presumably existed in earlier census data even though there was no way to explicitly identify them. See Section 2 on imputation for discussion of one difference between 2000 and earlier censuses.) Furthermore the analysis of A.C.E.II involves a number of special complications and assumptions beyond those of the standard dual system analyses.

## 2. Imputation

We use *II*, the Census Bureau's notation, to denote the number of imputations. Technically *II* is referred to as "insufficient information". It is not unusual for some census records to contain incomplete information to a modest extent. If all or nearly all relevant information is missing so that the matching of the P-sample records to the E-sample enumerations is not feasible, then the record is described as having insufficient information. Here we use the word "imputation" generally to describe records that for some reason do not include enough information to be included in the A.C.E. process. Broadly speaking, census imputation also includes imputation for item non-response for records in the A.C.E., and imputation for matching status in the A.C.E. process. Yet in our context, imputation is referred to as the whole records not included in the A.C.E. process due to insufficient information. In the 2000 census, imputations included two parts: inherent imputation and late adds.

One can identify two basic kinds of inherent imputation. Sometimes we do know with reasonable certainty how many people there are in the household, but lack personal information about them as is needed for the matching of the E-sample and the P-sample in the dual system process. In this case, we just need to impute demographic information for each person. On the other hand, sometimes the actual number of people in the household is also unknown. In this circumstance both the true counts and personal information need to be imputed. It is even possible to give a finer subdivision of types of inherent imputations. See Norwood and Citro [11].

Imputation related to a large number of late-adds was a special feature of the 2000 census. Because of its concern about address duplication, the Census Bureau created a special research program just after the basic census data was collected. The Bureau was able to identify, and pulled out, approximately 6 million person records in 2.4 million housing units as potential duplicates. Later on, approximately 2.4 million persons in 1 million housing units were reinstated into the census. However, this was too late for the 2.4 million people to be included in the A.C.E. process. Hence they were referred to as "Late Adds" and were treated similarly to imputation data. For details of research on duplicates, see ESCAP [4]. Table 1 is a comparison of the distributions of imputation in 1990 and 2000.

Besides the fact that there was no special treatment for Late Adds in the 1990 census, there is a significant difference in terms of the ratio of imputations from households with known person count and imputations from households with unknown person count between the 1990 and the 2000 Census. In 2000, that ratio was about 4. Yet in 1990, the ratio was 44 which is 10 times larger than that in 2000.

TABLE 1
*Number of imputations (II) as a percentage of census count (C)*

| Imputation type | 2000 Census | 1990 Census |
|---|---|---|
| Known Person Count | 1.68 | 0.88 |
| Unknown Person Count | 0.43 | 0.02 |
| Late Adds | 0.85 | 0.00 |
| Total | 2.96 | 0.90 |

(Source: *The 2000 Census: Interim Assessment*)

The percentages of $II$ from the 1980 census were more similar to those of 2000 than were the 1990 percentages.

In this paper, the item $C-II$ denotes the number of people with full information. They are frequently referred to as "data-defined" persons, and we use $DD$ to denote the number of them in the following sections.

## 3. Dual system estimation

As we introduced before, the 2000 A.C.E. data consists of the E-sample and the P-sample. Based on the information of the E-sample and the P-sample, a dual system estimate of the population is produced for special subgroups, called post-strata. These post-strata estimates are then apportioned and recombined so as to form estimates for any geographic area, such as state, county, census block etc. We now discuss some aspects of this procedure.

### 3.1. Post-stratification

For the purpose of analysis, the population is divided into certain groups called post-strata. Sixty-four post-stratum groups were created based on information about geographic location, race, Hispanic origin, housing tenure etc. In addition there were 7 age/sex categories. Thus originally there were 448 post-strata. Later on, some small post-strata were collapsed together to form 416 final post-strata. [See Table 5 in the Appendix for details of the construction of post-strata.]

### 3.2. Dual system estimation

The dual system estimate for post-stratum $i$ can be written as

$$(1) \qquad \widehat{DSE}_i = DD_i \times \widehat{CR}_i \times \frac{1}{\widehat{MR}_i}.$$

Here $DD_i$ is the number of data-defined persons in post-stratum $i$. $\widehat{CR}_i$ and $\widehat{MR}_i$ are the estimates of the E-sample correct enumeration rate and the P-sample matching rate respectively.

In the E-sample, enumerations are divided into two categories: correct enumerations and erroneous enumerations. The correct enumeration rate measures the accuracy of the census. It is estimated as

$$(2) \qquad \widehat{CR}_i = \frac{CE_i}{CE_i + EE_i},$$

where $CE_i$ denotes the number of correct enumerations and $EE_i$ denotes the number of erroneous enumerations in post-stratum $i$.

The P-sample persons are taken into a matching procedure to see whether they can be matched with persons in the E-sample. The P-sample matching rate then measures the coverage of the census. The formula for $\widehat{MR}_i$ is more complicated than that for the other elements of (1), and it is not particularly pertinent to the current considerations. The reader should consult Hogan [8] for details.

Since it was adopted by the Census Bureau to estimate the population, the dual system estimation method has been considered in principle a large-scale capture-recapture procedure. It can be motivated from an over-simplified, primitive model for capture-recapture estimation. In this model, the interrelation of the P-sample and the E-sample can be schematically summarized in a two by two table, and elements in the two by two table are estimated based on the assumption of the independence of the E-sample and the P-sample. For a detailed overview of dual system estimation, see Hogan [7].

### 3.3. Synthetic assumption

The census provides population figures for geographic subdivisions much smaller than those defined by post-stratum boundaries. These "smaller areas" include states, congressional districts, metropolitan areas, and even divisions as small as census tracts and census blocks within tracts.

In order to get smaller area estimates, the estimates $\widehat{DSE}_i$ for each post-stratum must be divided up and apportioned to geographic areas lying within that post-stratum. This procedure is called synthetic estimation and the assumption(s) that support its validity is (are) referred to as the synthetic assumption.

It seems to us that there are various reasonable forms of synthetic assumptions that could be proposed, and these lead in practice to different smaller area population estimates. For now we first present the formula implemented by the Bureau. Then we later contrast it with alternative formulas that also seem to us to be plausible.

For the purpose of synthetic estimation, the Census Bureau assumes that the estimate, $\widehat{DSE}_i$, should be divided in proportion to the total census counts within its post-stratum. Let the index $k$, $k = 1, 2, \ldots, K_i$ refer to geographic subregions within post-stratum $i$. Let $C_{ik}$ denote the total census counts for post-stratum $i$ and region $k$, and let $C_i$ denote the totals for the post-stratum. The Bureau population estimate for post-stratum $i$ region $k$ is then called $\widehat{DSE}_{ik}$ or $S_{ik}$ and is given by the formula

$$(3) \qquad S_{ik} \equiv \widehat{DSE}_{ik} = \frac{C_{ik}}{C_i} \widehat{DSE}_i.$$

This reflects the Bureau's synthetic assumption that the population distribution for smaller areas within a post-stratum is homogeneous with respect to the census counts for those areas within that post-stratum.

Formula (3) is often rephrased in a different but equivalent format. Define the Coverage Correction Factor for post-stratum $i$ ($CCF_i$) by

$$(4) \qquad CCF_i = \frac{\widehat{DSE}_i}{C_i}.$$

Then

$$(5) \qquad S_{ik} = C_{ik} CCF_i.$$

There is a different but equivalent way to interpret (3) or (5). The Census Bureau's estimate can also be written as

$$(6) \qquad S_{ik} = C_{ik} + (\widehat{DSE}_i - C_i) \times \frac{C_{ik}}{C_i}.$$

We will later build upon this interpretation.

In summary, for geographic region $k$ this gives the following population estimate:

$$(7) \qquad S_k = \sum_i S_{ik} = \sum_i C_{ik} CCF_i.$$

Here in (7), $S_k$ is called the synthetic dual system estimate, abbreviated as SynDSE. It is clear from its definition that it applies the same adjustment factor for people in each post-stratum, and aggregates the adjusted post-stratum level population numbers for an estimate of the population of the entire geographic area.

### 3.4. Rationale for post-stratification

The preceding discussion highlights one main rationale and target for post-stratification. Accuracy of the synthetic estimation formula (3) rests on the assumption that the population for the geographic areas within post-strata is distributed in proportion to the census count.

There are at least two other reasons for post-stratification in connection with dual system estimation. The logic supporting the dual system estimate requires that the matching rate be constant for individuals within post-strata. Violation of this will, in general, lead to bias in the dual system estimate (1) of the post-stratum population. Such a situation is referred to as "correlation bias". There are many discussions of correlation bias in the literature. For example, Seker and Deming [12] had an early discussion on correlation bias. Bell [1] introduced a third system to estimate the correlation bias. Freedman and Wachter [5] also had a discussion on correlation bias and heterogeneity. Zhao [14] investigated the data of the 2000 census to test the plausibility of the assumption of absence of correlation bias.

A third, though perhaps less important, rationale for post-stratification is that, in principle, suitably chosen post-strata can reduce the variance of estimates given through formulas such as (1) and (3). Conversely, a choice of too many post-strata with consequently small sample sizes within each post-stratum can lead to estimators with inflated variances. See Hogan [7] for a discussion of this in relation to the 1990 census. See Freedman and Wachter [6] for a perspective on post-stratification and its effects in the 2000 census.

## 4. Alternative formulas

In this section, we present three alternative formulas for synthetic estimation. The Census Bureau's formula is based on the synthetic assumption that the population distribution for small areas within a post-stratum is homogeneous with respect to the census counts (including imputations) for those areas within that post-stratum. Our alternative formulas are sensitive to the the homogeneity of imputations in the census, and its role in the synthetic estimation of subpopulation counts.

### 4.1. First alternative formula

Note that the estimates $\widehat{DSE}_i$ are computed only from enumerations of data-defined people. That is because $C_i$ does not appear in (1). Thus the estimates of $\widehat{DSE}_i$ of post-stratum totals involve $DD$ directly, but do not involve the number of counts labelled as $II$. It can thus be plausibly argued that the counts $II$ should also not play a role in distributing $\widehat{DSE}_i$ geographically within post-strata.

As noted in Section 3.4, homogeneity assumptions relative to the components of (1) are already part of the general justification for dual system estimation. From this perspective, it also seems reasonable to assume that the population for the geographic area within post-strata should be proportional to the enumeration of data defined people. This form of synthetic assumption leads to the alternate estimate $S_{ik}^1$ described as the formula

$$(8) \qquad S_{ik}^1 \equiv \widehat{DSE}_{ik}^1 = \frac{DD_{ik}}{DD_i} \widehat{DSE}_i,$$

where $DD_{ik}$ is the number of data-defined persons in geographic region $k$ within post-stratum $i$, $i = 1, 2, \ldots, I$, $k = 1, 2, \ldots, K_i$.

There is another way to view the formula for $S_{ik}^1$. For each post-stratum $i$, consider $DCF_i$ (Data-defined Coverage Factor) as a replacement of $CCF_i$. Their relationship is described in the following formula

$$(9) \qquad DCF_i = \frac{C_i}{C_i - II_i} = \frac{C_i}{DD_i} CCF_i.$$

Then applying the same Data-defined Coverage Factor for post-stratum $i$ to the number of data-defined persons in geographic region $k$ within post-stratum $i$, the corresponding $S_{ik}^1$ for geographic level $k$ is thus written as

$$(10) \qquad S_{ik}^1 = DD_{ik} DCF_i.$$

Note that (9) implies that $DCF_i = \frac{\widehat{DSE}_i}{DD_i}$, it is easy to show that (8) and (10) are equivalent.

### 4.2. Second alternative formula

It can be plausibly argued that the distribution of imputations $II_{ik} = C_{ik} - DD_{ik}$, $k = 1, 2, \ldots, K_i$ is a valid reflection of distribution of the true undercount relative to $C_{ik}$ within the post-stratum. Presumably imputations are concentrated in areas where it is intrinsically hard to count people, and hence areas with high undercount rate would be expected to have high imputation rate. Since the "true" undercount is not observed, it is hard, or impossible to devise a way to check this assertion. If it were valid, then the desirable estimate for the true population would be derived by distributing the post-stratum undercount estimates within the post-stratum in proportion to $II_{ik}$. This leads to the formula

$$(11) \qquad \begin{aligned} S_{ik}^2 &= C_{ik} + (\widehat{DSE}_i - C_i) \times \frac{II_{ik}}{II_i} \\ &= C_{ik} + (DD_i \times DCF_i - C_i) \times \frac{II_{ik}}{II_i}. \end{aligned}$$

As we noted before, the estimate of the total undercount for post-stratum $i$ is $\widehat{DSE}_i - C_i$, and this undercount is distributed to each geographic level proportionally to its imputation rate within the post-stratum. The estimate for the population is then the census counts plus the estimated undercount. In summary, this formula is the same as (6) except that $\frac{II_{ik}}{II_i}$ is substituted for $\frac{C_{ik}}{C_i}$.

### 4.3. Third alternative formula

Note that the Census Bureau's formula (6) is $S_k = C_{ik} + (\widehat{DSE}_i - C_i) \times \frac{C_{ik}}{C_i}$. Compare this with (11), and another reasonable formula comes out naturally as

$$
\begin{aligned}
(12) \quad S_{ik}^3 &= C_{ik} + (\widehat{DSE}_i - C_i) \times \frac{DD_{ik}}{DD_i} \\
&= C_{ik} + (DD_i \times DCF_i - C_i) \times \frac{DD_{ik}}{DD_i}.
\end{aligned}
$$

In words, this formula begins from a base of the census counts $C_{ik}$ (including $II_{ik}$). It then considers the distribution of $DD_{ik}$ as a reflection of the true undercount rate at geographic level within post-strata.

Clearly all of the formulas presented here have the same normalization property

$$
(13) \qquad \sum_k S_{ik}^l = \sum_k S_{ik} = \widehat{DSE}_i, \qquad l = 1, 2, 3.
$$

Also, if we take the summation over post-stratum index $i$, then we will have the estimate of the population at geographic area $k$ as

$$
(14) \qquad S_k^l = \sum_i S_{ik}^l, \qquad l = 1, 2, 3.
$$

## 5. Results from alternative formulas at state level

### 5.1. Comparison of shares at state level

Allocating seats in the House of Representatives is the original constitutional mandate for which the decennial census was established. Much attention was put on which states had gained or lost seats. It is of primary interest to compare different formulas at the state level.

Figure 1 shows comparison of alternative formulas and the Census Bureau's formula for the 16 largest states. [See Figure 5 in the Appendix for the full comparison of all 51 states.] The comparison is made in the sense of population shares. A state's population share is normally defined as its percentage of the national total. Thus they do not affect estimates for national totals. The horizontal line for each state shows the confidence interval of share difference: SynDSE ($S_k$) share minus census share. The standard error of share difference is computed from Davis [3] published by the Census Bureau. The square represents the share difference between $S_k$ and census, the dot represents the share difference between $S_k^1$ and census, and the triangle represents the share difference between $S_k^2$ and census. The share difference between $S_k^3$ and census is omitted from the figure since it is very close to the one between $S_k$ and census.

The most prominent feature is for the state of New York where the difference calculated from $S_k^1$ falls very far outside of (below) the confidence interval calculated from census formula. For several other states the result for $S_k^1$ is also outside the confidence interval (above, as for North Carolina, Virginia, and Ohio, or below, as for Indiana and Illinois). $S_k^2$ agrees better with the census formula. For several large states, such as Texas, California, Florida and Pennsylvania, the square and the triangle are very close to each other. The result for New York is driven towards 0, although it still falls outside (above) the confidence interval.

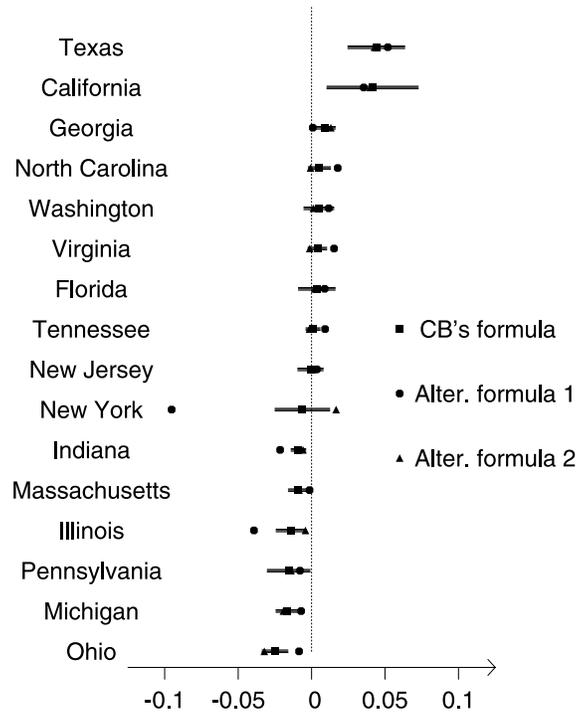

FIG 1. *State level shares comparison from different formulas.*

Interestingly, most of the time, the share difference of $S_k$ and census falls between the difference of $S_k^1$ and census, and the difference of $S_k^2$ and census. This tells us that, in a sense the census formula is a compromise of the two alternatives we introduced.

### 5.2. Role of imputation

Imputations create the primary difference in practice between the Bureau's synthetic formula (3) and alternative formulas such as our (8), (11) and (12). Note that the assumption justifying (8) is that the undercount is homogeneous with respect to $DD_{ik}$ for regions within post-strata. In contrast, the assumption justifying (3) is that of homogeneity with respect to $C_{ik} = DD_{ik} + II_{ik}$. If the imputation rates were stochastically homogeneous with respect to $C_{ik}$, then both formulas would have the same expectation, and would generally yield very similar results in practice.

Imputation rates for the 16 large states of Figure 1, together with the population shares from the census, are given in Table 2. In this table, the total imputation rates, the imputation rates from late adds (LA) and non late adds (Non-LA), as well as the census shares are listed. [See Table 6 in the Appendix for the full table for all 51 states.] The overall imputation rate for New York is considerably larger than the national rate of 3%.

Furthermore, what really matters is the imputation rates within post-strata within the state relative to those post-strata results elsewhere. Because of this it seems informative to supplement the overall imputation rates given in the table

| State | II(Tot) | II(Non-LA) | II(LA) | Census Share | Number of post-strata | Mean II(Tot) of post-strata |
|-------|---------|------------|--------|--------------|----------------------|-----------------------------|
| NY | 4.913 | 3.201 | 1.712 | 6.724 | 256 | 5.511 |
| TX | 3.508 | 2.633 | 0.875 | 7.417 | 264 | 3.962 |
| IL | 3.383 | 2.469 | 0.914 | 4.422 | 246 | 4.380 |
| GA | 3.300 | 2.349 | 0.951 | 2.907 | 234 | 3.994 |
| CA | 3.255 | 2.720 | 0.535 | 12.08 | 260 | 4.360 |
| NJ | 2.869 | 2.008 | 0.861 | 3.004 | 194 | 4.849 |
| NC | 2.795 | 1.640 | 1.156 | 2.849 | 236 | 3.558 |
| IN | 2.700 | 2.202 | 0.498 | 2.157 | 246 | 4.106 |
| FL | 2.672 | 2.113 | 0.558 | 5.700 | 236 | 4.534 |
| MA | 2.468 | 1.558 | 0.909 | 2.240 | 253 | 3.239 |
| TN | 2.465 | 1.599 | 0.867 | 2.025 | 222 | 3.372 |
| WA | 2.407 | 1.894 | 0.513 | 2.105 | 236 | 3.410 |
| PA | 2.322 | 1.574 | 0.748 | 4.331 | 242 | 3.595 |
| VA | 2.283 | 1.555 | 0.727 | 2.503 | 260 | 3.536 |
| MI | 1.876 | 1.341 | 0.536 | 3.541 | 260 | 3.039 |
| OH | 1.680 | 1.123 | 0.557 | 4.040 | 222 | 2.699 |

with per post-strata averages. As a result, Table 2 also gives the mean imputation rate per post-strata within state as computed from the following formula:

$$(15) \qquad MIR_k = \frac{1}{n_k^*} \sum_{\{i, C_{ik} \neq 0\}} \frac{II_{ik}}{C_{ik}} \times 100\%$$

where $n_k^*$ is the number of post-strata within the state with non-zero census counts, which is also listed in the table. Even a cursory examination of these imputation rates in the census reveals that an assumption for the imputations of stochastic homogeneity within post-strata is not reasonable. (A valid, formal test of this statistical hypothesis can be derived using the methods of Zhao [14]. This test decisively rejects the null hypothesis of stochastic homogeneity, with a *p*-value < 0.0001.)

In Table 2, the comparison of New York and New Jersey points to an interesting phenomenon. Overall New Jersey has an imputation rate of 2.869%. This is fairly close to the national average. But it shares a lot of post-strata with New York. The mean value of the imputation rates per post-strata in New Jersey is 4.849%. This is the second highest among the 16 states. Yet as shown in Figure 1, in contrast to New York, the differences for New Jersey using $S_k^1$ and $S_k^2$ are quite close to that using the Census Bureau's $S_k$. The result is that although New Jersey has relatively high mean imputation rate per post-strata, its population estimate is not increased as much by the dual system as this might seem to warrant. One explanation for this is that an important neighboring state (New York) has even higher imputation rates.

From another point of view, we can consider our alternative formula one as a basic rate for estimate of population, while the Census Bureau's formula can be viewed as an attempt to use imputations with the hope of improving these basic estimates.

## 6. Results from alternative formulas at county-group level

To better investigate the differences among all the formulas, we conduct a further analysis down to a finer level: county-group level. Ideally our analysis might have been performed on the level of congressional districts. However we had only county level data to work with. Hence we created county groups to roughly approximate the size and geographic contiguity of congressional districts. (In some cases our county groups were much more populous than congressional districts since we could not split counties into smaller districts.) In general, small adjacent counties are lumped to form a group with population roughly like a congressional district, while relatively large counties (for example, a county contains several congressional districts) would make a county-group by themselves. Totally we created 369 county-groups, on average each having 730,000 people.

For each county-group, an adjusted estimate ($SynDSE$) is constructed by the Census Bureau's formula and our alternative formula 1, 2 and 3. It seems most suitable to compare the adjustments to the relative shares. This is consistent with the discussion in Brown et al. [2] and Freedman and Wachter [5]. However we found direct statements of share differences to be less suitable in part because of unfamiliarity with the county-groups and variability in their sizes. Hence it seems more informative to express the adjustments in percentage terms from a base of the original census numbers. It can be easily shown that this measure is a linear transformation of the share difference, and as noted in the above references, the results from the percent adjustment would be consistently comparable to the share difference.

There are two possible choices of the base of the original census numbers. Naturally people would consider the census counts, and the relative percent difference can be expressed as

$$(16) \qquad reldif^c = \frac{SynDSE - C}{C} \times 100\%.$$

However, one of the implications of the alternative formula one is that the number of data-defined person $DD$ is a more basic quantity. Therefore we use $DD$ as the base, and the relative percent difference is then defined as

$$(17) \qquad reldif^d = \frac{SynDSE - DD}{DD} \times 100\%.$$

To account for the implication of imputation, (17) can be modified to be a measure called state adjusted difference ($SAD$), which is defined by

$$(18) \qquad SAD_j = (\frac{SynDSE_j - DD_j}{DD_j} - \frac{II_s}{DD_s}) \times 100\%.$$

In (18), $j$ is the county-group index, $s$ is the state index. The following Table 3 illustrates the descriptive statistics for $SAD$ using different formulas.

As we already found from the last section, the alternative formula three gives very similar results as the Census Bureau's. It is also noticeable from the table that overall there is no substantial difference in terms of the mean value of differences. [The results from $reldif^c$ can be found in Table 7 in the Appendix, and they will give similar relative conclusions among county groups within a state.]

TABLE 3
*Distribution of state adjusted difference at county group level*

|                | Min   | Max  | Median | Mean | SD   |
|----------------|-------|------|--------|------|------|
| CB's formula   | −2.97 | 7.38 | 0.98   | 1.14 | 1.40 |
| Alter. formula 1 | −2.95 | 4.93 | 1.20   | 1.18 | 1.10 |
| Alter. formula 2 | −3.08 | 9.80 | 0.81   | 1.15 | 1.60 |
| Alter. formula 3 | −2.98 | 7.29 | 0.99   | 1.14 | 1.39 |

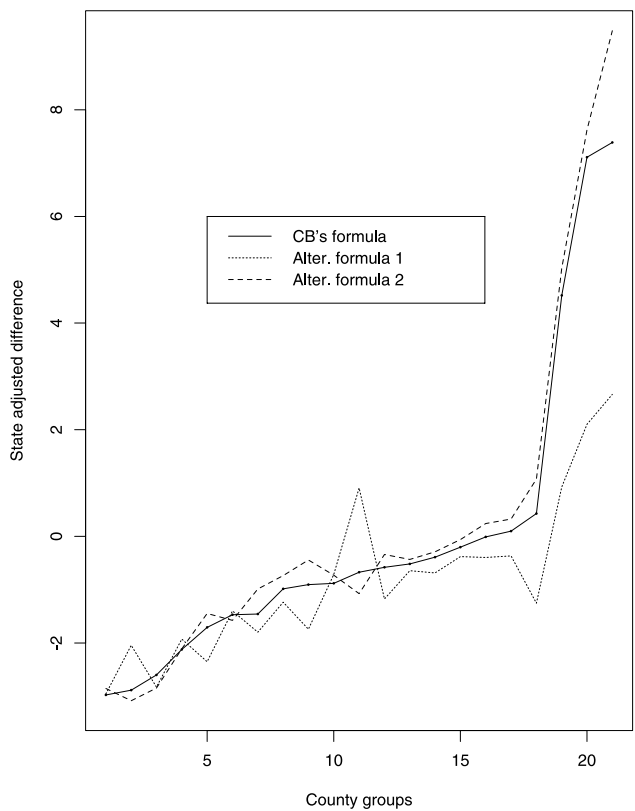

FIG 2. *State adjusted difference – New York (DD base).*

It is impossible to visually show the results of $SAD$ from all county-groups in one figure; instead we illustrate the results in the following three states:

1. New York: because of the large discrepancy in share comparison (Figure 1) and the relatively large size (3rd biggest state)
2. New Jersey: because of the interesting phenomenon discussed in Section 5.2
3. California: because of the relatively large size (biggest state)

Figure 2 is the plot of $SAD$ in each county-group in New York. [The table generating this figure can be found in the Appendix.] Each one of the 21 points on the X-axis represents a county group, and the state adjusted differences represented on the Y-axis are connected by a line. Different types of lines represent different formulas. Again, the results from alternative formula 3 are not shown in the figure because they are very close to those from the Census Bureau's formula. It is obvious that for the three counties in New York city (Bronx, Kings and Queens) which have a very large percent of imputation, the differences are much higher than those from other county-groups.

Figure 3 is the plot of $SAD$ in each county-group in New Jersey. Despite the fact that New Jersey shares a lot of post-strata with New York, the scale of the differences is much smaller than that from New York.

Figure 4 is the plot of $SAD$ in each county-group in California. From all three figures, it can be seen that most of the time, the lines using Census Bureau's formula lie between the lines using our alternative formula 1 and alternative formula 2.

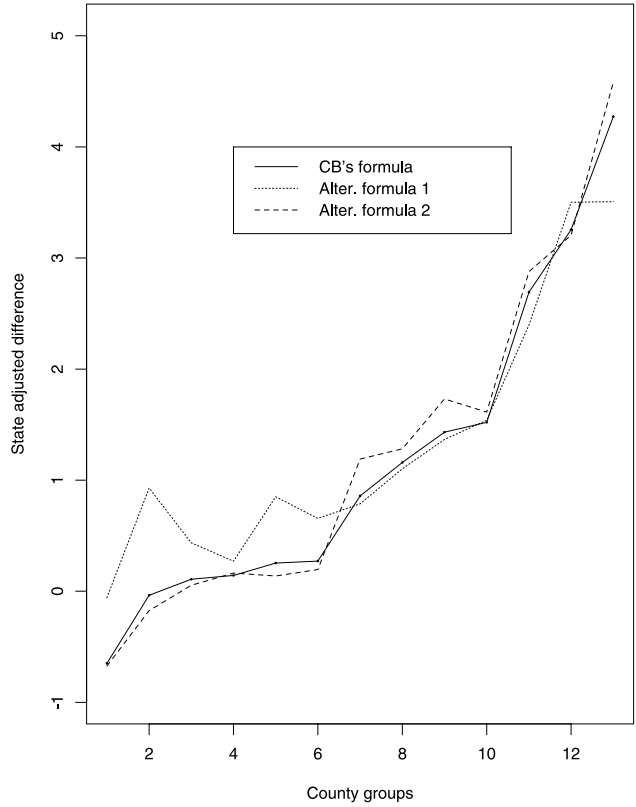

Fig 3. *State adjusted difference – New Jersey (DD base).*

This confirms that the Census Bureau's formula is kind of a compromise of the two alternatives.

It can also be seen that in general, at the lower end of the figure (smaller difference between $SynDSE$ and $DD$), the difference using Census Bureau's formula tends to be lower (higher) than that using alternative formula 1 (using alternative formula 2), while at the upper end of the figure (larger difference between $SynDSE$ and $DD$), the difference using Census Bureau's formula tends to be higher (lower) than that using alternative formula 1 (using alternative formula 2). (The detailed results at each county group in these three states could be found in Table 8 through Table 10 in the Appendix.)

## 7. Comparison of different formulas

### 7.1. Comparison of four formulas

As stated earlier, if the imputation rates were stochastically homogeneous with respect to the census count, then all the formulas would have the same expectation. It is easy to prove that if $\frac{II_{ik}}{II_i} = \frac{C_{ik}}{C_i}$, then $S_k^1 = S_k^2 = S_k^3 = S_k$.

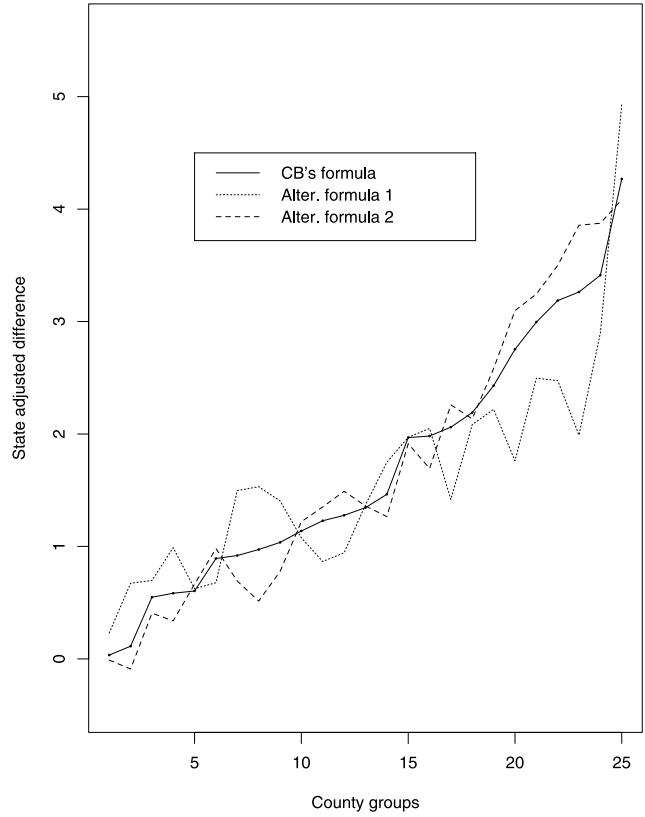

Fig 4. *State adjusted difference – California (DD base).*

### 7.2. When is DCF better

Our alternative formula (10) uses $DCF$ instead of $CCF$. One may wonder under which conditions does $DCF$ behave better than $CCF$.

Consider the following simpler case: there are two states for a single post-stratum, and there are no people who moved between the census day and the A.C.E. interview. The corresponding counts in state 1 and 2 within post-stratum are: $CE_1, CE_2, EE_1, EE_2, MN_1, MN_2, NN_1, NN_2, II_1, II_2$, and they are all observable. Here $CE_j$, $EE_j$, $MN_j$, $NN_j$, and $II_j$ $(j = 1, 2)$ denotes the number of correct enumerations, erroneous enumerations, matched non-movers, unmatched non-movers, and imputations respectively. For a formal definition of these types of counts, see Norwood and Citro [11]. As also shown in Norwood and Citro [11], $CCF$ and $DCF$ can be written as functions of these five types of counts

$$(19) \qquad CCF = \frac{CE_1 + CE_2}{CE_1 + CE_2 + EE_1 + EE_2 + II_1 + II_2} \times \frac{NN_1 + NN_2}{MN_1 + MN_2},$$

$$(20) \qquad DCF = \frac{CE_1 + CE_2}{CE_1 + CE_2 + EE_1 + EE_2} \times \frac{NN_1 + NN_2}{MN_1 + MN_2}.$$

To further simplify the case, we assume that the two states are equal in size, i.e. $CE_1 = CE_2, MN_1 = MN_2, NN_1 = NN_2$

The following analysis makes a comparison of the squared errors resulting from use of (3) and (10). In order to make this comparison it is necessary to make some assumptions about the true population. The analysis is somewhat simple under the plausible assumption that the unbiased DSE from the two by two tables within each state describes the true population parameters. A similar analysis is possible under other assumptions.

The unbiased DSE from the actual two by two tables within each state can be written as

$$S_1^t = CE_1 \times \frac{NN_1}{MN_1} = S, \qquad S_2^t = CE_2 \times \frac{NN_2}{MN_2} = S.$$

The synthetic DSEs for state 1 and 2 within post-stratum calculated from $CCF$ and $DCF$ (use alternative formula one) are

$$(21) \qquad S_i^c = \frac{CE_i + EE_i + II_i}{CE_1 + CE_2 + EE_1 + EE_2 + II_1 + II_2} \times 2S, \qquad i = 1, 2,$$

and

$$(22) \qquad S_i^d = \frac{CE_i + EE_i}{CE_1 + CE_2 + EE_1 + EE_2} \times 2S, \qquad i = 1, 2.$$

Define the variance, i.e. the squared error of synthetic DSE from the true population, as

$$\begin{aligned}
\Delta_c &= (S_1^c - S_1^t)^2 + (S_2^c - S_2^t)^2 \\
&= 2S^2 \left( \frac{CE_1 + EE_1 + II_1 - (CE_2 + EE_2 + II_2)}{CE_1 + EE_1 + II_1 + CE_2 + EE_2 + II_2} \right)^2, \\
\Delta_d &= (S_1^d - S_1^t)^2 + (S_2^d - S_2^t)^2 \\
&= 2S^2 \left( \frac{CE_1 + EE_1 - (CE_2 + EE_2)}{CE_1 + EE_1 + CE_2 + EE_2} \right)^2.
\end{aligned}$$

The difference of $\Delta_d$ and $\Delta_c$ is

$$\begin{aligned}
(23) \\
\Delta_d - \Delta_c &= 2S^2 \left\{ \left( \frac{EE_1 - EE_2}{2CE_1 + EE_1 + EE_2} \right)^2 \right. \\
&\qquad \left. - \left( \frac{EE_1 + II_1 - (EE_2 + II_2)}{2CE_1 + EE_1 + EE_2 + II_1 + II_2} \right)^2 \right\} \\
&= 2S^2 \left\{ \left( \frac{EE_1 - EE_2}{2CE_1 + EE_1 + EE_2} + \frac{EE_1 + II_1 - EE_2 - II_2}{2CE_1 + EE_1 + EE_2 + II_1 + II_2} \right) \right. \\
&\qquad \left. \left( \frac{EE_1 - EE_2}{2CE_1 + EE_1 + EE_2} - \frac{EE_1 + II_1 - EE_2 - II_2}{2CE_1 + EE_1 + EE_2 + II_1 + II_2} \right) \right\}.
\end{aligned}$$

If $CE >> (EE, II)$, as is usually the case, then

$$(24) \\
\Delta_d - \Delta_c \approx \frac{-2S^2 (4CE_1(EE_1 - EE_2) + 2CE_1(II_1 - II_2))(2CE_1(II_1 - II_2))}{(2CE_1 + EE_1 + EE_2)^2 (2CE_1 + EE_1 + EE_2 + II_1 + II_2)^2}.$$

TABLE 4
*Frequency table of better performance of DCF among large/small post-strata*

|  | **CCF** | **DCF** | **Total** |
|---|---|---|---|
| Small | 48 | 89 | 137 |
| Large | 23 | 84 | 107 |
| Total | 71 | 173 | 244 |

From (24) we have

- If $EE_1 = EE_2$ then $\Delta_d - \Delta_c \leq 0$, $DCF$ is better.
- If $II_1 = II_2$ then $\Delta_d - \Delta_c \geq 0$, $CCF$ is better.
- If $EE_1 \neq EE_2$ and $II_1 \neq II_2$.
  - If $EE_1 > EE_2$ and $II_1 > II_2$ then $\Delta_d - \Delta_c \leq 0$, $DCF$ is better.
  - If $EE_1 > EE_2$ and $II_1 < II_2$.
    * If $EE_1 - EE_2 \leq -\dfrac{II_1 - II_2}{2}$ then $\Delta_d - \Delta_c \leq 0$, $DCF$ is better.
    * If $EE_1 - EE_2 > -\dfrac{II_1 - II_2}{2}$ then $\Delta_d - \Delta_c > 0$, $CCF$ is better.

More generally, we assume $CE_2 = \lambda CE_1$, $MN_2 = \lambda MN_1$, $NN_2 = \lambda NN_1$, since homogeneity assumption appears to hold for the two largest groups: CE and MN. For the setup and results from the test of homogeneity assumption, see Zhao [14]. Similarly we have

- If $\lambda EE_1 = EE_2$ then $\Delta_d - \Delta_c \leq 0$, $DCF$ is better.
- If $\lambda II_1 = II_2$ then $\Delta_d - \Delta_c \geq 0$, $CCF$ is better.
- If $\lambda EE_1 \neq EE_2$ and $\lambda II_1 \neq II_2$.
  - If $\lambda EE_1 > EE_2$ and $\lambda II_1 > II_2$ then $\Delta_d - \Delta_c \leq 0$, $DCF$ is better.
  - If $\lambda EE_1 > EE_2$ and $\lambda II_1 < II_2$.
    * If $\lambda EE_1 - EE_2 \leq -\dfrac{\lambda II_1 - II_2}{2}$ then $\Delta_d - \Delta_c \leq 0$, $DCF$ is better.
    * If $\lambda EE_1 - EE_2 > -\dfrac{\lambda II_1 - II_2}{2}$ then $\Delta_d - \Delta_c > 0$, $CCF$ is better.

The above discussion gives certain conditions when the Census Bureau's correction factor (4) or the alternative correction factor (9) performs better than the other one. To show the empirical results from the data, let's consider a simple case. Suppose we regard New York state as state 1, and all the other states together as state 2, then we calculate the $DCF$ and $CCF$ for the 244 post-strata that are in both states. We found that $DCF$ is better in 70% of post-strata which exist in both state 1 and state 2. Furthermore, if we categorize the post-strata into two groups: large post-strata (having more than 50,000 correct enumerations) and small post-strata, $DCF$ performs much better in the large post-strata.

From Table 4, it could be seen that $DCF$ (corresponding to formula (10)) performs better about 65% of the time in small post-strata and 80% of time in large post-strata.

## 8. Conclusion

The major purpose of this paper is to better understand the 2000 A.C.E. process by providing alternative formulas. To construct these three formulas, alternate forms of the synthetic assumption are used, and the structure of imputation is analyzed. We find that the alternative estimation formulas seem also justifiable.

It is perhaps hard to tell which formula gives generally more accurate results. It appears to us that each one has its own merit and no one dominates another. In addition, there seems no way with existing data to compare the biases of the formulas. Nonetheless, it appears that the first of the alternatives would achieve smaller variance than that of the Census Bureau's formula if the number of erroneous enumerations and the number of imputations are positively correlated, which holds true in most of the cases.

What we do observe is that the Census Bureau's formula tends to be a compromise among the three alternatives. For this reason it seems to us reasonable to stick to the original one, especially in view of a lack of further evidence.

All the Census Bureau's formula and our alternative formulas use the total number of imputations to create population estimates. As noted in Section 2, there are different classes of imputation. It may be preferable to use only some subsets of imputations, and create formulas in different ways.

Finally we want to point out that the correct enumeration rate $CE/(CE + EE)$ is estimated in producing synthetic estimation. This estimate is another potential source of heterogeneity, and the related synthetic assumption on it should be studied. A valid, formal test of the hypothesis that the correct enumeration rate is geographically homogeneous within post-strata for states or counties can be derived using the methods of Zhao [14]. This test shows there is significant non-homogeneity. (The details of this test will be reported elsewhere.) It would be desirable to also see how this inhomogeneity affects synthetic estimates results. However, unlike $II$, the components $CE$ and $EE$ are not measured for the entire census, but rather only for the A.C.E. sample blocks. Thus it is unclear how to use existing data to create estimates related to this factor.

## Appendix

TABLE 5. *Schematic for post-stratification variables (see Section 3.1 for further description) (MSA: Metropolitan Statistical Area; TEA: Type of Enumeration Area; MO/MB: Mail out/Mail back)*

| Race/Hispanic Origin / Domain number | Tenure | MSA/TEA | High return rate | | | | Low return rate | | | |
|---|---|---|---|---|---|---|---|---|---|---|
| | | | NE | MW | S | W | NE | MW | S | W |
| Domain 7: Non-Hispanic White and Other | Owner | Large MSA MO/MB | 1 | 2 | 3 | 4 | 5 | 6 | 7 | 8 |
| | | Medium MSA MO/MB | 9 | 10 | 11 | 12 | 13 | 14 | 15 | 16 |
| | | Small MSA & Non-MSA MO/MB | 17 | 18 | 19 | 20 | 21 | 22 | 23 | 24 |
| | | All Other TEAs | 25 | 26 | 27 | 28 | 29 | 30 | 31 | 32 |
| | Non-Owner | Large MSA MO/MB | | 33 | | | | 34 | | |
| | | Medium MSA MO/MB | | 35 | | | | 36 | | |
| | | Small MSA & Non-MSA MO/MB | | 37 | | | | 38 | | |
| | | All Other TEAs | | 39 | | | | 40 | | |
| Domain 4: Non-Hispanic Black | Owner | Large MSA MO/MB | | 41 | | | | 42 | | |
| | | Medium MSA MO/MB | | | | | | | | |
| | | Small MSA & Non-MSA MO/MB | | 43 | | | | 44 | | |
| | | All Other TEAs | | | | | | | | |
| | Non-Owner | Large MSA MO/MB | | 45 | | | | 46@ | | |
| | | Medium MSA MO/MB | | | | | | | | |
| | | Small MSA & Non-MSA MO/MB | | 47 | | | | 48 | | |
| | | All Other TEAs | | | | | | | | |
| Domain 5: Native Hawaiian or Pacific Islander | | Owner | | | | 49 | | | | |
| | | Non-Owner | | | | 50 | | | | |
| Domain 6: Non-Hispanic Asian | | Owner | | | | 51 | | | | |
| | | Non-Owner | | | | 52 | | | | |
| Domain 3: Hispanic | Owner | Large MSA MO/MB | | 53 | | | | 54 | | |
| | | Medium MSA MO/MB | | | | | | | | |
| | | Small MSA & Non-MSA MO/MB | | 55 | | | | 56 | | |
| | | All Other TEAs | | | | | | | | |
| | Non-Owner | Large MSA MO/MB | | 57 | | | | 58 | | |
| | | Medium MSA MO/MB | | | | | | | | |
| | | Small MSA & Non-MSA MO/MB | | 59 | | | | 60 | | |
| | | All Other TEAs | | | | | | | | |
| Domain 1: On Reservation American Indian or Alaska Native | | Owner | | | | 61 | | | | |
| | | Non-Owner | | | | 62 | | | | |
| Domain 6: Off Reservation American Indian or Alaska Native | | Owner | | | | 63 | | | | |
| | | Non-Owner | | | | 64 | | | | |

 *L. Brown and Z. Zhao*

TABLE 6
*Imputation rates for 51 states*

| State | II(Tot) | II(Non-LA) | II(LA) | Census Share | Number of post-strata | Mean II(Tot) of post-strata |
|-------|---------|------------|--------|--------------|-----------------------|------------------------------|
| NY | 4.913 | 3.201 | 1.712 | 6.724 | 256 | 5.511 |
| NM | 4.474 | 2.895 | 1.579 | 0.652 | 236 | 4.137 |
| HI | 4.247 | 2.913 | 1.334 | 0.430 | 222 | 4.781 |
| WY | 3.921 | 2.588 | 1.333 | 0.175 | 148 | 4.166 |
| NV | 3.918 | 3.257 | 0.661 | 0.718 | 236 | 4.479 |
| AZ | 3.891 | 3.145 | 0.746 | 1.835 | 236 | 4.942 |
| VT | 3.887 | 2.223 | 1.664 | 0.215 | 134 | 3.961 |
| DC | 3.860 | 3.726 | 0.134 | 0.196 | 144 | 4.373 |
| TX | 3.508 | 2.633 | 0.875 | 7.417 | 264 | 3.962 |
| AL | 3.491 | 2.212 | 1.279 | 1.584 | 235 | 4.104 |
| IL | 3.383 | 2.469 | 0.914 | 4.422 | 246 | 4.380 |
| DE | 3.339 | 2.901 | 0.438 | 0.277 | 222 | 4.900 |
| RI | 3.302 | 2.360 | 0.942 | 0.369 | 221 | 4.821 |
| GA | 3.300 | 2.349 | 0.951 | 2.907 | 234 | 3.994 |
| CA | 3.255 | 2.720 | 0.535 | 12.08 | 260 | 4.360 |
| SC | 3.221 | 2.145 | 1.076 | 1.417 | 236 | 4.423 |
| MD | 3.074 | 2.503 | 0.572 | 1.887 | 222 | 4.178 |
| NH | 3.056 | 1.987 | 1.070 | 0.439 | 217 | 4.440 |
| MT | 3.039 | 1.583 | 1.456 | 0.321 | 152 | 4.478 |
| MS | 3.038 | 1.677 | 1.360 | 1.005 | 228 | 3.289 |
| LA | 2.886 | 1.886 | 1.000 | 1.584 | 236 | 3.432 |
| NJ | 2.869 | 2.008 | 0.861 | 3.004 | 194 | 4.849 |
| AR | 2.810 | 1.403 | 1.407 | 0.950 | 222 | 2.852 |
| NC | 2.795 | 1.640 | 1.156 | 2.849 | 236 | 3.558 |
| CO | 2.786 | 2.039 | 0.747 | 1.535 | 236 | 4.119 |
| IN | 2.700 | 2.202 | 0.498 | 2.157 | 246 | 4.106 |
| FL | 2.672 | 2.113 | 0.558 | 5.700 | 236 | 4.534 |
| ME | 2.604 | 1.258 | 1.345 | 0.453 | 184 | 2.889 |
| AK | 2.584 | 1.385 | 1.199 | 0.202 | 152 | 3.131 |
| ID | 2.554 | 1.821 | 0.733 | 0.461 | 152 | 3.374 |
| WV | 2.506 | 0.856 | 1.651 | 0.645 | 215 | 2.625 |
| MA | 2.468 | 1.558 | 0.909 | 2.240 | 253 | 3.239 |
| TN | 2.465 | 1.599 | 0.867 | 2.025 | 222 | 3.372 |
| KT | 2.447 | 1.164 | 1.283 | 1.435 | 222 | 2.888 |
| WA | 2.407 | 1.894 | 0.513 | 2.105 | 236 | 3.410 |
| CT | 2.390 | 1.544 | 0.847 | 1.205 | 256 | 3.875 |
| UT | 2.369 | 1.765 | 0.604 | 0.801 | 236 | 2.972 |
| SD | 2.362 | 1.392 | 0.970 | 0.266 | 140 | 2.733 |
| PA | 2.322 | 1.574 | 0.748 | 4.331 | 242 | 3.595 |
| VA | 2.283 | 1.555 | 0.727 | 2.503 | 260 | 3.536 |
| OK | 2.261 | 1.282 | 0.979 | 1.220 | 236 | 2.517 |
| OR | 2.260 | 1.711 | 0.549 | 1.222 | 236 | 3.639 |
| WI | 2.153 | 1.600 | 0.553 | 1.903 | 258 | 4.311 |
| MO | 2.098 | 1.200 | 0.898 | 1.986 | 222 | 3.050 |
| ND | 1.985 | 1.003 | 0.983 | 0.226 | 138 | 2.342 |
| KS | 1.904 | 1.227 | 0.678 | 0.953 | 236 | 2.739 |
| MI | 1.876 | 1.341 | 0.536 | 3.541 | 260 | 3.039 |
| MN | 1.873 | 1.237 | 0.636 | 1.748 | 236 | 3.463 |
| OH | 1.680 | 1.123 | 0.557 | 4.040 | 222 | 2.699 |
| IA | 1.629 | 0.963 | 0.666 | 1.032 | 215 | 2.589 |
| NE | 1.608 | 0.994 | 0.615 | 0.607 | 236 | 2.487 |

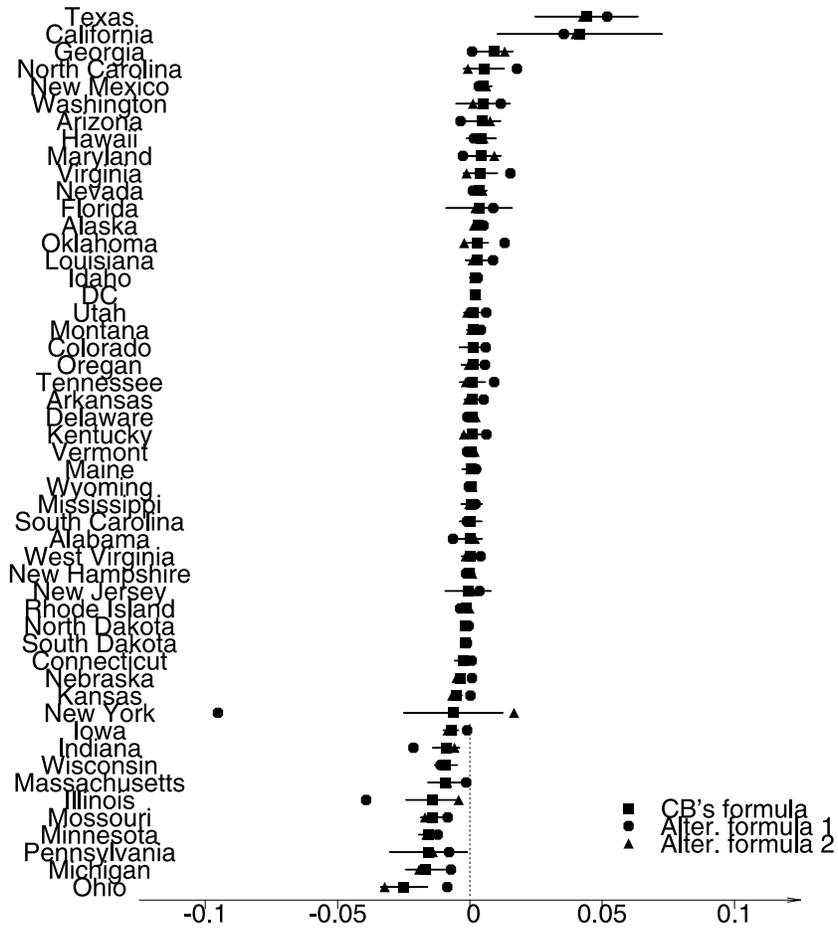

Fig 5. *Share comparison at state level.*

Table 7

*Distribution of relative difference between <u>census</u> and <u>SynDSE</u> at county group level*

| | Min | Max | Median | Mean | SD |
|---|---|---|---|---|---|
| CB's formula | −0.13 | 2.96 | 1.15 | 1.14 | 0.54 |
| Alter. formula 1 | −3.26 | 3.78 | 1.22 | 1.18 | 0.84 |
| Alter. formula 2 | −0.14 | 4.31 | 1.06 | 1.13 | 0.68 |
| Alter. formula 3 | −0.13 | 2.97 | 1.15 | 1.14 | 0.54 |

Table 8

*County group level results in New Jersey*
*(Through Table 8 to 10, the second column "CB's" lists the results using the Census Bureau's formula, the third column "Alter. 1" lists the results using alternative formula 1, and the fourth column "Alter. 2" lists the results using alternative formula 2.)*

| Relative difference in New Jersey (census as the base) | | | | | |
|---|---|---|---|---|---|
| Counties | CB's | Alter. 1 | Alter. 2 | Census | II/Census |
| Passaic | 1.566 | 1.282 | 1.743 | 479073 | 3.863 |
| Essex | 1.471 | 1.706 | 1.423 | 770844 | 4.462 |
| Hudson | 1.470 | 0.745 | 1.763 | 599525 | 5.369 |
| Somerset, Union | 1.250 | 1.268 | 1.339 | 807714 | 3.087 |
| Atlantic, Cape May & Cumberland, Salem | 1.234 | 1.179 | 1.354 | 542964 | 2.766 |
| Mercer | 1.223 | 1.160 | 1.514 | 329669 | 3.030 |
| Middlesex, Monmouth | 1.123 | 1.707 | 1.009 | 1334607 | 2.021 |
| Morris | 1.064 | 1.387 | 1.011 | 461026 | 1.938 |
| Sussex, Warren | 0.973 | 1.920 | 0.838 | 243450 | 1.890 |
| Bergen | 0.967 | 1.344 | 0.894 | 872769 | 2.187 |
| Burlington, Ocean | 0.964 | 1.089 | 0.985 | 912247 | 2.068 |
| Camden, Gloucester | 0.951 | 0.882 | 1.274 | 747998 | 2.756 |
| Hunterdon | 0.797 | 1.380 | 0.772 | 117643 | 1.474 |

| State adjusted difference in New Jersey | | | | | |
|---|---|---|---|---|---|
| Counties | CB's | Alter. 1 | Alter. 2 | DD | II/DD |
| Hudson | 4.273 | 3.507 | 4.583 | 567337 | 5.674 |
| Essex | 3.255 | 3.502 | 3.206 | 736452 | 4.670 |
| Passaic | 2.693 | 2.398 | 2.877 | 460565 | 4.019 |
| Somerset, Union | 1.521 | 1.540 | 1.613 | 782780 | 3.185 |
| Mercer | 1.432 | 1.367 | 1.732 | 319680 | 3.125 |
| Atlantic, Cape May & Cumberland, Salem | 1.159 | 1.103 | 1.282 | 527948 | 2.844 |
| Camden, Gloucester | 0.858 | 0.787 | 1.190 | 727384 | 2.834 |
| Bergen | 0.271 | 0.656 | 0.196 | 853681 | 2.236 |
| Middlesex, Monmouth | 0.254 | 0.851 | 0.138 | 1307639 | 2.062 |
| Burlington, Ocean | 0.143 | 0.270 | 0.164 | 893380 | 2.112 |
| Morris | 0.108 | 0.437 | 0.054 | 452090 | 1.977 |
| Sussex, Warren | −0.036 | 0.930 | −0.174 | 238849 | 1.926 |
| Hunterdon | −0.649 | −0.057 | −0.674 | 115909 | 1.496 |

Table 9
*County group level results in New York*

### Relative difference in New York (census as the base)

| Counties | CB's | Alter. 1 | Alter. 2 | Census | II/Census |
|---|---|---|---|---|---|
| Bronx | 2.405 | −1.893 | 4.313 | 1285415 | 9.016 |
| Clinton, Franklin, Fulton, Hamilton & | | | | | |
| Jefferson , Lewis Oswego, St Lawrence | 1.683 | 1.512 | 1.819 | 524735 | 3.126 |
| Chenango, Delaware , Herkimer & | | | | | |
| Madison, Oneida, Otsego, Schoharie | 1.486 | 1.364 | 1.568 | 530826 | 3.023 |
| Broome, Sullivan, Tioga, Tompkins, Ulster | 1.480 | 1.193 | 1.576 | 562836 | 3.147 |
| New York | 1.477 | 3.012 | 1.088 | 1477358 | 2.887 |
| Allegany, Cattaraugus, Chautauqua & | | | | | |
| Chemung, Schuyler, Steuben, Yates | 1.280 | 1.036 | 1.522 | 484489 | 2.786 |
| Dutchess, Putnam | 1.263 | 0.891 | 1.506 | 355568 | 3.702 |
| Kings | 1.255 | −3.263 | 1.717 | 2426027 | 9.817 |
| Orange, Rockland | 1.212 | 0.764 | 1.430 | 606779 | 3.851 |
| Columbia, Essex, Greene, Rensselaer & | | | | | |
| Saratoga, Warren, Washington | 1.095 | 0.520 | 1.326 | 603542 | 3.339 |
| Westchester | 0.999 | 1.162 | 1.149 | 899806 | 3.150 |
| Albany, Montgomery, Schenectady | 0.999 | 1.067 | 0.896 | 469399 | 2.602 |
| Queens | 0.945 | −2.359 | 1.417 | 2202506 | 7.967 |
| Cayuga, Cortland, Onondaga | 0.794 | 0.986 | 0.807 | 567471 | 2.184 |
| Nassau | 0.736 | −0.072 | 1.183 | 1312886 | 3.380 |
| Monroe | 0.735 | 1.566 | 0.541 | 708834 | 1.512 |
| Erie | 0.684 | 0.057 | 0.939 | 919474 | 2.683 |
| Niagara, Orleans | 0.532 | 0.311 | 0.291 | 256313 | 1.984 |
| Suffolk | 0.491 | 0.161 | 0.947 | 1390791 | 3.103 |
| Genesee, Livingston, Ontario & | | | | | |
| Seneca, Wayne, Wyoming | 0.252 | 0.273 | 0.372 | 376399 | 1.899 |
| Richmond | −0.035 | −1.626 | 0.576 | 434542 | 5.332 |

### State adjusted difference in New York

| Counties | CB's | Alter. 1 | Alter. 2 | DD | II/DD |
|---|---|---|---|---|---|
| Bronx | 7.386 | 2.662 | 9.482 | 1169523 | 9.909 |
| Kings | 7.110 | 2.100 | 7.622 | 2187875 | 10.885 |
| Queens | 4.517 | 0.927 | 5.030 | 2027022 | 8.657 |
| Richmond | 0.428 | −1.252 | 1.073 | 411372 | 5.632 |
| Orange, Rockland | 0.098 | −0.367 | 0.325 | 583412 | 4.005 |
| Dutchess, Putnam | −0.011 | −0.397 | 0.241 | 342405 | 3.844 |
| Clinton, Franklin, Fulton, Hamilton & | | | | | |
| Jefferson, Lewis Oswego, St Lawrence | −0.203 | −0.379 | −0.062 | 508331 | 3.227 |
| Broome, Sullivan, Tioga, Tompkins, Ulster | −0.390 | −0.686 | −0.291 | 545126 | 3.249 |
| Chenango, Delaware, Herkimer & | | | | | |
| Madison, Oneida, Otsego, Schoharie | −0.517 | −0.643 | −0.433 | 514779 | 3.117 |
| Columbia, Essex, Greene, Rensselaer & | | | | | |
| Saratoga Warren, Washington | −0.580 | −1.175 | −0.340 | 583388 | 3.455 |
| New York | −0.673 | 0.908 | −1.073 | 1434701 | 2.973 |
| Westchester | −0.882 | −0.714 | −0.728 | 871460 | 3.253 |
| Nassau | −0.906 | −1.742 | −0.443 | 1268496 | 3.499 |
| Allegany, Cattaraugus, Chautauqua & | | | | | |
| Chemung, Schuyler, Steuben, Yates | −0.985 | −1.236 | −0.736 | 470993 | 2.865 |
| Suffolk | −1.457 | −1.798 | −0.987 | 1347631 | 3.203 |
| Albany, Montgomery, Schenectady | −1.470 | −1.400 | −1.575 | 457183 | 2.672 |
| Erie | −1.708 | −2.351 | −1.445 | 894808 | 2.757 |
| Cayuga, Cortland, Onondaga | −2.123 | −1.927 | −2.109 | 555079 | 2.232 |
| Niagara, Orleans | −2.601 | −2.826 | −2.846 | 251228 | 2.024 |
| Monroe | −2.885 | −2.042 | −3.082 | 698113 | 1.536 |
| Genesee, Livingston, Ontario & | | | | | |
| Seneca, Wayne, Wyoming | −2.974 | −2.953 | −2.852 | 369250 | 1.930 |

TABLE 10
*County group level results in California*

| Relative difference in California (census as the base) | | | | | |
|---|---|---|---|---|---|
| **Counties** | **CB's** | **Alter. 1** | **Alter. 2** | **Census** | **II/Census** |
| Imperial | 2.959 | 3.589 | 2.783 | 131317 | 4.343 |
| Kings | 2.384 | 1.891 | 2.827 | 109332 | 4.114 |
| San Luis Obispo, Santa Barbara | 2.191 | 2.257 | 1.911 | 613840 | 2.995 |
| Monterey, San Benito, Santa Cruz | 2.086 | 1.607 | 2.325 | 680087 | 4.018 |
| Merced, Stanislaus | 2.050 | 1.848 | 2.198 | 647207 | 3.538 |
| Del Norte, Humboldt, Lake, Mendocino, Napa | 1.918 | 1.920 | 1.865 | 406509 | 3.242 |
| Kern, Tulare | 1.913 | 1.233 | 2.213 | 993655 | 4.352 |
| Los Angeles | 1.829 | 1.727 | 1.776 | 9344086 | 3.529 |
| Butte, Lassen, Modoc, Nevada, Plumas & Shasta, Sierra Siskiyou, Trinity, Yuba | 1.801 | 2.345 | 1.353 | 621777 | 2.431 |
| Fresno, Madera, Mariposa | 1.661 | 0.448 | 2.228 | 912453 | 4.657 |
| Colusa, Glenn, Sutter, Tehama, Yolo | 1.577 | 1.935 | 1.325 | 338148 | 2.704 |
| San Francisco | 1.572 | 0.623 | 1.900 | 756976 | 4.283 |
| Inyo,San Bernardino | 1.542 | 1.817 | 1.349 | 1682190 | 3.135 |
| Alameda | 1.408 | 1.969 | 1.186 | 1416006 | 2.757 |
| San Joaquin | 1.390 | 0.771 | 1.581 | 544827 | 3.827 |
| Riverside | 1.380 | 1.404 | 1.395 | 1511034 | 3.179 |
| Santa Clara | 1.282 | 1.223 | 1.361 | 1652871 | 3.081 |
| Orange | 1.275 | 0.953 | 1.483 | 2803924 | 3.216 |
| San Diego | 1.228 | 1.623 | 0.989 | 2716820 | 2.616 |
| San Mateo | 1.192 | 0.841 | 1.311 | 696711 | 3.252 |
| Ventura | 1.131 | 1.150 | 1.189 | 739985 | 2.729 |
| Sacramento | 1.105 | 1.250 | 0.966 | 1198004 | 2.702 |
| Contra Costa, Solano | 1.065 | 1.612 | 0.868 | 1316047 | 2.330 |
| Alpine, Amador, Calaveras, El Dorado & Mono, Placer, Tuolumne | 0.987 | 0.778 | 1.073 | 534773 | 3.136 |
| Marin, Sonoma | 0.953 | 1.144 | 0.909 | 683315 | 2.365 |

| State adjusted difference in California | | | | | |
|---|---|---|---|---|---|
| **Counties** | **CB's** | **Alter. 1** | **Alter. 2** | **DD** | **II/DD** |
| Imperial | 4.269 | 4.928 | 4.085 | 125614 | 4.540 |
| Kings | 3.412 | 2.898 | 3.874 | 104834 | 4.291 |
| Fresno, Madera, Mariposa | 3.262 | 1.990 | 3.857 | 869960 | 4.884 |
| Kern, Tulare | 3.186 | 2.475 | 3.500 | 950411 | 4.550 |
| Monterey, San Benito, Santa Cruz | 2.995 | 2.496 | 3.244 | 652762 | 4.186 |
| San Francisco | 2.753 | 1.762 | 3.090 | 724551 | 4.475 |
| Merced, Stanislaus | 2.429 | 2.219 | 2.582 | 624309 | 3.668 |
| Los Angeles | 2.189 | 2.084 | 2.135 | 9014370 | 3.658 |
| San Joaquin | 2.060 | 1.417 | 2.259 | 523974 | 3.980 |
| Sanluis Obispo, Santa Barbara | 1.982 | 2.049 | 1.693 | 595458 | 3.087 |
| Del Norte, Humboldt, Lake, Mendocino, Napa | 1.968 | 1.970 | 1.913 | 393332 | 3.350 |
| Inyo, San Bernardino | 1.464 | 1.748 | 1.265 | 1629458 | 3.236 |
| Riverside | 1.344 | 1.369 | 1.360 | 1462999 | 3.283 |
| Orange | 1.276 | 0.944 | 1.491 | 2713751 | 3.323 |
| San Mateo | 1.228 | 0.866 | 1.351 | 674056 | 3.361 |
| Santa Clara | 1.137 | 1.076 | 1.219 | 1601952 | 3.179 |
| Colusa, Glenn, Sutter, Tehama, Yolo | 1.036 | 1.404 | 0.777 | 329003 | 2.780 |
| Butte, Lassen, Modoc, Nevada, Plumas & Shasta, Sierra, Siskiyou, Trinity, Yuba | 0.972 | 1.530 | 0.513 | 606664 | 2.491 |
| Alameda | 0.920 | 1.496 | 0.691 | 1376961 | 2.836 |
| Alpine, Amador, Calaveras, El Dorado & Mono, Placer, Tuolumne | 0.892 | 0.677 | 0.981 | 518002 | 3.238 |
| Ventura | 0.604 | 0.623 | 0.664 | 719791 | 2.806 |
| San Diego | 0.584 | 0.988 | 0.337 | 2645741 | 2.687 |
| Sacramento | 0.548 | 0.698 | 0.407 | 1165633 | 2.777 |
| Contra Costa, Solano | 0.112 | 0.672 | −0.090 | 1285381 | 2.386 |
| Marin, Sonoma | 0.034 | 0.230 | −0.011 | 667156 | 2.422 |